\documentstyle[amsmath,amssymb,epsfig]{mn}

\title{A neutral hydrogen distance limit to the relativistic binary PSR~J1141$-$6545 } 
\author[S. M. Ord, M. Bailes and W. van Straten]
       {S.~M.~Ord,$^1$ 
       M.~Bailes$^1$ and
	W.~van~Straten$^1$\\
$^1$Swinburne University of Technology, \\
        Centre for Astrophysics and Supercomputing \\
        Mail 31 \\
        P. O. Box 218 \\
        VIC 3122 \\
        Australia}

\pagerange{\pageref{firstpage}--\pageref{lastpage}}
\pubyear{2002}

\begin{document} 
\maketitle 

\bibliographystyle{mn}
\begin{abstract}

We have obtained an HI absorption spectrum of the
relativistic binary PSR~J1141$-$6545 and used it to constrain the distance to the system.  The spectrum suggests that the pulsar is
at, or beyond, the tangent point, estimated to be at 3.7 kpc.  PSR~J1141$-$6545
offers the promise of stringent tests of General Relativity (GR) by comparing its
observed orbital period derivative with that derived from other relativistic
observables.  At the distance of PSR~J1141$-$6545 it should be possible to
verify GR to an accuracy of just a few percent,
as contributions to the observed orbital period derivative from kinematic terms
will be a small fraction of that induced by the emission of gravitational
radiation.  PSR~J1141$-$6545 will thus make an exceptional gravitational
laboratory.

\end{abstract} 
\nokeywords
\begin{keywords}
pulsars:general -- pulsars:individual (PSR~J1141~-~6545) -- stars:distances
\end{keywords}

\section{Introduction} 

An independent confirmation of General Relativity (GR) has recently been obtained
using a measurement of the annual orbital parallax of the binary millisecond
pulsar PSR~J0437$-$4715 to derive the inclination angle of the binary, thus
predicting the shape of the Shapiro delay \cite{vbb+01}. In this test, as with
those using self-consistency checks, GR has been shown to be a
remarkably accurate description of gravity.  Measurement of post-Keplerian
($PK$) orbital parameters in relativistic binary pulsars \cite{dt92} have
provided the most stringent tests of GR to date.  PSR~B1913+16 displays a measurable
orbital period derivative ($\dot{P_b}$), time dilation and gravitational
redshift parameter ($\gamma$) and an advance of periastron ($\dot{\omega}$),
which are all consistent with GR \cite{tw89,dt91,tay94}.
Unfortunately, at the $\sim$1\% level, these tests cannot be authoritative
because of our ignorance of the distance and exact contribution of the Galactic
potential to the observed orbital period derivative of the binary system.
PSR~B1534+12 is a nearby relativistic pulsar which demonstrates all of the
above $PK$ phenomena, together with the range ($r$) and shape ($s$) of Shapiro
delay which have been shown to be remarkably self-consistent with
GR \cite{sac+98}. As predicted by Bell and Bailes (1996)\nocite{bb96}, this pulsar could not
be used to verify the emission of gravitational radiation at better than the
15\% level because of the uncertain distance to the pulsar. Instead, Stairs et al.~(1998)
have used GR to obtain the pulsar distance to a high degree of accuracy.

The relativistic binary pulsar PSR~J1141$-$6545 was recently
discovered in the Parkes multibeam pulsar survey \cite{klm+00a}. It has a spin
period ($P$) of 394~ms, a very narrow pulse ($0.01P$) and is in an eccentric,
4.7~hr orbit. The measured $\dot{\omega}$ is 5.3 degrees~yr$^{-1}$ and it has a
predicted $\dot{P_b}$ of $-3.85 \times 10^{-13}$, which is twice that observed
for PSR~B1534+12, but nearly an order of magnitude less than that observed for
PSR~B1913+16. Tests of GR using PSR~J1141$-$6545 would nicely complement those
already made using the other relativistic systems because it is
thought to contain a neutron star and a white dwarf, rather than two neutron stars, as
in the PSR~B1534+12 and PSR~B1913+16 systems. Within a few years, the orientation and
component masses should be known to high accuracy from measurements of the
advance of periastron and $\gamma$ terms. This will completely specify the
orbital geometry of the system and make a specific prediction about the
magnitude of the orbital period derivative. 

A dispersion measure ($DM$) of 116 pc~cm$^{-3}$ places this pulsar at a distance
of 3.2 kpc using the Taylor and Cordes (1993)\nocite{tc93} model.  These
estimates can be significantly in errorr. In individual cases the error in the
DM--distance can be as great as a factor of 2. If this distance is correct the
kinematic ``contamination'' of the observed orbital period derivative would be
small.  Hence, PSR~J1141$-$6545 could be an extremely good gravitational
laboratory.


 The detection of neutral hydrogen absorption features in the spectrum of a
pulsar, together with a detailed neutral hydrogen emission spectrum in the same
direction, can be used to place constraints upon its distance.  A discussion of
the neutral hydrogen distance determination method is given by Frail and
Weisberg (1990)\nocite{fw90} and the most recent review is by Weisberg
(1996)\nocite{wei96}. In this paper we use neutral hydrogen measurements to
constrain the distance of PSR J1141$-$6545 and show that it should be an
exceptional laboratory for testing GR.

\section{Observations and method} 

The observations were carried out over 3 days from the 11th of January to the
13th of January 2001 using the Parkes 64\,m radio telescope.  The observing
system used was the Caltech-Parkes-Swinburne recorder (CPSR) \cite{vbb00}; a
baseband recorder which performs 2-bit sampling on two 20-MHz bands, recording
the data onto four parallel DLT 7000 tapes for off--line processing.  The
observations were centred in frequency at 1413 MHz, which places neutral
hydrogen (1420.406 MHz) comfortably in the observing band. The observations
were reduced at Swinburne University of Technology on a Beowulf-style cluster
of workstations.


In order to examine the structure of the neutral hydrogen absorption and
emission features in the pulsar spectrum we employed a software filterbank
using standard Fourier techniques.  16384 channels were formed across the
20~MHz observing band. The Stokes I parameter was obtained, by summation of the
detected left and right linear polarisations, and then folded at the
topocentric period of the pulsar. Due to the high frequency resolution, the
time resolution was reduced to 0.8192 ms. Folded profiles with 256 phase bins
were formed and individual integrations were summed together using an accurate
pulsar ephemeris to produce a single archive consisting of approximately 15.3
hours of data.  In order to increase the signal to noise ratio in an individual
channel the frequency resolution was reduced by a factor of 7 to 8.5 kHz. This
corresponds to a velocity resolution of 1.8~km~s$^{-1}$.

The emission spectrum was formed by calculating the mean level in the off-pulse
region and the pulsar spectrum was formed by subtracting the off-pulse
spectrum from the on-pulse spectrum.  The resultant pulsar spectrum was not
completely flat due to interstellar scintillation effects, but nevertheless, two
major absorption features are clearly visible in it. 
The error bars presented in the absorption spectrum were calculated
from the root mean square deviation of the off-pulse region of the profile in
each frequency channel.  This was a direct measure of the error associated with
each point in the absorption spectrum.  

In transforming the wavelength range into a velocity range account was made of
the orbital velocity of the Earth and the velocity of the solar system relative
to the local dynamical standard of rest (LSR). The variation of the  rotational
velocity of the Earth, an effect of maximum magnitude 0.4~km~s$^{-1}$ for a pulsar at
this declination ($-65^{\circ}$) was neglected, as was the 0.15~km~s$^{-1}$
change in the Earth's orbital velocity during the observations. These contributions are
small compared to our presented velocity resolution of 1.8~km~s$^{-1}$.

\subsection{Calibration considerations}

As a software filterbank was employed, instrumental factors, such as differential
gain between channels, which plague analogue filter bank spectrography were not
an issue. The software filterbank had uniform response over the entire
bandpass. We did not, therefore, employ the technique of frequency switching to
ensure instrumental effects were removed in the production of emission spectra.
We calibrated the baseline subtracted emission spectrum from the neutral
hydrogen survey performed by Kerr et al. (1986)\nocite{kbjk86}. Following
Koriabalski et al. (1995)\nocite{kjww95} we consider the error in the peak
value of the brightness temperature to be 5~K. The structure of the emission
spectrum is in good agreement with that presented by Kerr et al. (1986).

The absorption spectrum is shown in Fig. \ref{spectrum} as both a ratio of the
measured absorption to the baseline, and as an optical depth, $\tau$: where $I
= I_0 \times e^{-\tau}$. 

\section{Distance determination}

\begin{figure}
\psfig{file = 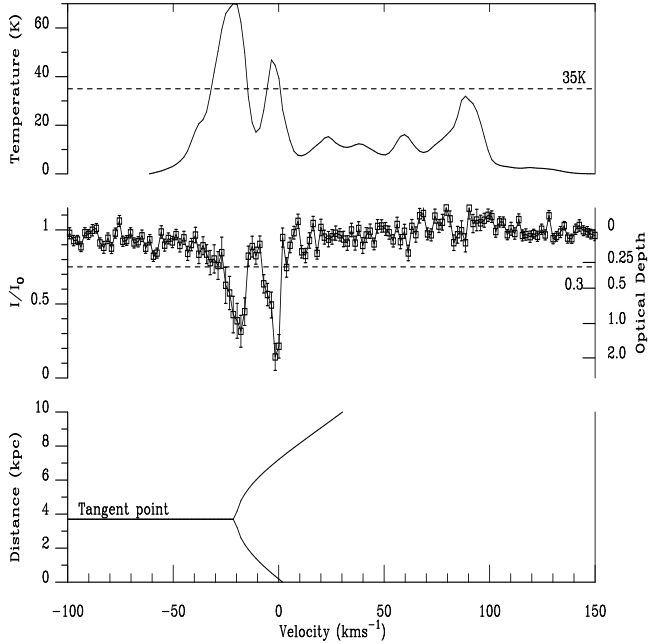,width=8.5cm,height=8.5cm,angle=270}
\caption{The upper panel is the neutral hydrogen emission spectrum in the
	direction of PSR~J1141$-$6545, the middle panel is the absorption
		spectrum of the pulsar and the bottom panel is the
		Galactic
		rotation model of Fich, Blitz and Stark evaluated in the
		direction of the pulsar. The dashed lines represent
		significance levels. If light from the pulsar has passed
		through an emission feature stronger than 35K, corresponding
		absorption features should be present in the pulsar spectrum
		with an optical depth greater than 0.3.} \label{spectrum}
\end{figure}

In order to ascribe a distance to any feature in either the absorption or
emission spectrum we transformed the observed velocity into a distance via a
Galactic rotation model. The model used was that of Fich, Blitz \& Stark
(1989)\nocite{fbs89}. A systematic uncertainty of 7~km~s$^{-1}$ is assumed in the
velocity--distance conversion.  Even with this error it is clear from Fig.
\ref{spectrum} that there is significant emission at velocities forbidden by
the rotation model.

The Galactic coordinates of J1141-6545 are {\it l}=295$^{\circ}$.77, {\it
b}=$-$3$^{\circ}$.8. This is in the direction of the Carina spiral arm.
The emission spectrum clearly shows a peak at approximately
20~km~s$^{-1}$ which is associated with this spiral arm. Other
pulsars in this direction also display similar emission at forbidden
velocities. It is generally considered that the discrepancy between the model
and the observations in this direction is due to the streaming motion of the
gas in the Carina spiral arm  \cite{jkww96}.

A lower distance limit is found by ascertaining the distance of the most
distant absorption feature in the pulsar spectrum.  Upper distance limits are obtained
by a determination of the furthest significant feature in the neutral hydrogen
emission spectrum which does not have a corresponding absorption feature. The
relative strengths of the emission and absorption features are used to
determine not only the significance of the spectral features but can be used to
determine the temperature of the neutral hydrogen.

\begin{figure}
\psfig{file = 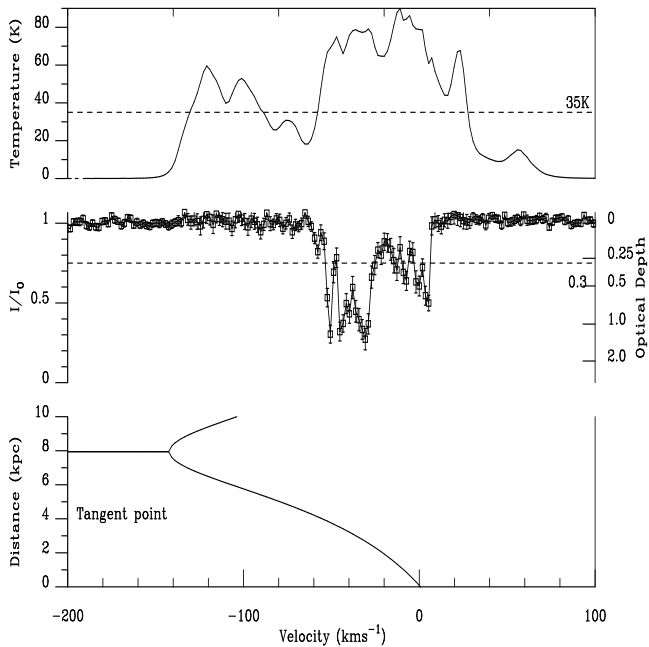,width=8.5cm,height=8.5cm,angle=270}
\caption{In order to test the techniques employed, we have observed the pulsar
	PSR~J1644$-$4559. The spectra are presented above, the panels and dashed lines are as described in
		Figure 1. The PSR~J1644$-$4559 spectra are in good agreement
		with those in the literature, and show no evidence of any spurious absrption features.} \label{1641}
\end{figure}

\subsection{Interpretation of features}

The important points to consider here are the significance of any
absorption feature and the likelihood of a given emission feature
appearing in the absorption spectrum.  In order to ascribe a level of
significance to absorption and emission features we applied the
criterion of Weisberg, Boriakoff and Rankin (1979)\nocite{wbr79}. That
is, any emission feature with brightness temperature ($T_B$), greater
than 35~K rarely displays an optical depth ($\tau$), less than
$0.3$. In applying this criterion to the spectra it is clear that
although the absorption is not associated with the emission all the
way out to the the velocity of the most negative emission feature,
there is significant absorption associated with all emission brighter
than 35~K. From this we ascribe a lower distance limit to this pulsar
of 3.7~kpc which is the distance of the tangent point in this
direction using the Fich, Blitz and Stark (1989) model. There is probably a
10 to 15 \% discrepancy between this distance estimation and the
actual distance to the tangent point \cite{fbs89}.

Obtaining an upper distance limit is not possible as there is no emission
feature at positive velocities greater than our chosen 35~K significance 
level.  

\subsubsection{Contamination of the Absorption Spectrum}

This type of observation is prone to contamination of the absorption spectrum
by features in the emission spectrum. This arises as the absorption spectrum is
formed by subtracting two profiles of similar amplitude. Any non-linearities in the 
digitisation or detection process will appear in the absorption spectrum as inverted
emission spectrum features. In order to ascertain the level of this
contamination, the observational technique applied to PSR~J1141$-$6545 was also
applied to PSR~J1644$-$4559, a bright pulsar with a well known neutral hydrogen
spectrum. The absorption spectrum for PSR~J1644$-$4559 is presented in Fig
\ref{1641} and is found to be in excellent agreement with that published in the
literature (Ables and Manchester 1976\nocite{am76}; Frail et al.
1991\nocite{fchw91}), which validates our experimental
technique. 
We thus consider the presented absorption spectrum of PSR~J1141$-$6545 to
be free of any appreciable contamination from features in the emission spectrum.  

\section{Discussion} 

The distance ascribed to this pulsar from its dispersion measure is 3.2 kpc.
The limit obtained here serves as an independent distance measure and allows a
number of consequences to be explored.

\subsection{Electron Density}

As the DM of the pulsar is known, we can use the neutral
hydrogen distance to give an upper limit on the electron density in this
direction. The DM of PSR~J1141$-$6545 is  approximately 116.2 pc cm$^{-3}$. A
distance limit of greater than 3.7 kpc provides an upper limit to the mean
electron density along this line of sight of 0.03 cm$^{-3}$.

\subsection{The Binary Orbital Period Derivative} 

The secular decrease of binary orbital period, $\dot{P_b}$, has been
successfully observed in PSR~B1913+16 \cite{tw89,dt91,tay94} and PSR~B1534+12
\cite{sac+98}. While the measured value is dominated by the emission of
gravitational radiation, unfortunately there are many possible contributions to
an observed orbital period derivative \cite{dt91}. They include;
the component predicted by GR; the
acceleration of the centre of mass of the binary system with respect to the Sun
due to the gravitational field of the Galaxy, and the apparent acceleration
induced by the proper motion of the system.  

Tauris and Sennels (2000)\nocite{ts00} have predicted that the space velocity
of the binary should be greater than 150~km~s$^{-1}$ but Kaspi et al.
(2000)\nocite{klm+00a} point out that the age and Galactic latitude of the
system suggest its velocity perpendicular to plane must be $\sim 150(d/3.2kpc)$
km~s$^{-1}$. It is possible to examine the magnitude of the kinematical
contribution to any measured orbital period derivative. For transverse
velocities in the range 100 - 200~km~s$^{-1}$ the level of the kinematical
contribution to $\dot{P_b}$ is just a few percent, but without a measurement of
the proper motion, we will be unable to confirm GR.  The prospects of making a
quick proper motion measurement are not clear. Young pulsars like
PSR~J1141$-$6545 have intrinsic timing noise which make proper motion
measurements difficult from timing on short baselines. Interferometric
observations may be more useful. PSR~J1141$-$6545 has a very narrow pulse and a
reasonable flux density at 20~cm wavelength. If a suitable reference could be
found, an accurate proper motion should be obtainable via VLBI within a few
years. 

An upper limit to the distance of PSR~J1141$-$6545 will
be more difficult to obtain. The distance method employed
by Stairs et al. (1998)\nocite{sac+98} for PSR~B1534$+$12 will be
less accurate in the case of PSR~J1141$-$6545, because the
kinematic terms are a much smaller fraction of the
relativistic contribution to the observed orbital
period derivative.

\section{Conclusion}

The measurement of the absorption spectrum of this pulsar has allowed
a lower limit to be placed upon its distance of 3.7 kpc, the
tangent point distance predicted by the Galactic rotation curve. 
The mean electron density in the direction of this pulsar is
therefore at most 0.03 cm$^{-3}$.
We have also examined the Galactic and kinematic contribution to the
observed $\dot{P_b}$ and found that it should only 
be at the few percent level for
reasonable pulsar transverse velocities making PSR~J1141$-$6545
an excellent gravitational laboratory. Future
measurements of the system's proper motion will
help to further define the level at which this
system can be used to test GR.


\begin{thebibliography}{{Weisberg, Boriakoff \& Rankin }{1979}}

\bibitem[\protect\citename{Ables \& Manchester }{1976}]{am76}
Ables~J.~G., Manchester~R.~N., 1976, Astr.\,Astrophys., 50, 177

\bibitem[\protect\citename{Bell \& Bailes }{1996}]{bb96}
Bell~J.~F., Bailes~M., 1996, Astrophys.\,J., 456, L33

\bibitem[\protect\citename{Damour \& Taylor }{1991}]{dt91}
Damour~T., Taylor~J.~H., 1991, Astrophys.\,J., 366, 501

\bibitem[\protect\citename{Damour \& Taylor }{1992}]{dt92}
Damour~T., Taylor~J.~H., 1992, 45, 1840

\bibitem[\protect\citename{Fich, Blitz \& Stark }{1989}]{fbs89}
Fich~M., Blitz~L., Stark~A.~A., 1989, Astrophys.\,J., 342, 272

\bibitem[\protect\citename{Frail \& Weisberg }{1990}]{fw90}
Frail~D.~A., Weisberg~J.~M., 1990, Astron.\,J., 100, 743

\bibitem[\protect\citename{Frail {\rm et~al. }}{1991}]{fchw91}
Frail~D.~A., Cordes~J.~M., Hankins~T.~H., Weisberg~J.~M., 1991, Astrophys.\,J.,
  382, 168

\bibitem[\protect\citename{Johnston {\rm et~al. }}{1996}]{jkww96}
Johnston~S., Koribalski~B.~S., Weisberg~J., Wilson~W., 1996,
  Mon.\,Not.\,R.\,astr.\,Soc., 279, 661

\bibitem[\protect\citename{Kaspi {\rm et~al. }}{2000}]{klm+00a}
Kaspi~V.~M. {\rm et~al.}, 2000, Astrophys.\,J., 543, 321

\bibitem[\protect\citename{Kerr {\rm et~al. }}{1986}]{kbjk86}
Kerr~F.~J., Bowers~P.~F., Jackson~P.~D., Kerr~M., 1986,
  Astr.\,Astrophys.\,Suppl.\,Ser., 66, 373

\bibitem[\protect\citename{Koribalski {\rm et~al. }}{1995}]{kjww95}
Koribalski~B.~S., Johnston~S., Weisberg~J., Wilson~W., 1995, Astrophys.\,J.,
  441, 756

\bibitem[\protect\citename{Stairs {\rm et~al. }}{1998}]{sac+98}
Stairs~I.~H., Arzoumanian~Z., Camilo~F., Lyne~A.~G., Nice~D.~J., Taylor~J.~H.,
  Thorsett~S.~E., Wolszczan~A., 1998, Astrophys.\,J., 505, 352

\bibitem[\protect\citename{{Tauris} \& {Sennels} }{2000}]{ts00}
{Tauris}~T.~M., {Sennels}~T., 2000, Astr.\,Astrophys., 355, 236

\bibitem[\protect\citename{Taylor \& Cordes }{1993}]{tc93}
Taylor~J.~H., Cordes~J.~M., 1993, Astrophys.\,J., 411, 674

\bibitem[\protect\citename{Taylor \& Weisberg }{1989}]{tw89}
Taylor~J.~H., Weisberg~J.~M., 1989, Astrophys.\,J., 345, 434

\bibitem[\protect\citename{Taylor }{1994}]{tay94}
Taylor~J.~H., 1994, 66, 711

\bibitem[\protect\citename{van Straten, Britton \& Bailes }{2000}]{vbb00}
van Straten~W., Britton~M., Bailes~M., 2000, in Kramer~M., Wex~N.,
  Wielebinski~R., eds, Pulsar Astronomy - 2000 and Beyond, {IAU} Colloquium
  177.
\newblock Astronomical Society of the Pacific, San Francisco, p.~283

\bibitem[\protect\citename{van Straten {\rm et~al. }}{2001}]{vbb+01}
van Straten~W., Bailes~M., Britton~M., Kulkarni~S.~R., Anderson~S.~B.,
  Manchester~R.~N., Sarkissian~J., 2001, Nature, 412, 158

\bibitem[\protect\citename{Weisberg }{1996}]{wei96}
Weisberg~J.~M., 1996, in Johnston~S., Walker~M.~A., Bailes~M., eds, Pulsars:
  Problems and Progress, {IAU} Colloquium 160.
\newblock Astronomical Society of the Pacific, San Francisco, p.~447

\bibitem[\protect\citename{Weisberg, Boriakoff \& Rankin }{1979}]{wbr79}
Weisberg~J.~M., Boriakoff~V., Rankin~J., 1979, Astr.\,Astrophys., 77, 204

\end{thebibliography}
\end{document}